\documentstyle[prl,epsf,eqsecnum,aps,floats]{revtex}  
\begin{document}
\def\theequation{\arabic{equation}}%
\def\dofig#1{\vskip.2in\centerline{\epsfbox{#1}}}
\newcommand{\rstev}{\mbox{$\rs = \T{1.8}$}}
\newcommand{\XX}{\mbox{$\, \times \,$}}
\newcommand{\AP}{\mbox{${\rm \bar{p}}$}}
\newcommand{\SU}{\mbox{$S$}}
\newcommand{\SPt}{\mbox{$<\! |S|^2 \!>$}}
\newcommand{\ET}{\mbox{$E_{T}$}}
\newcommand{\PT}{\mbox{$p_{t}$}}
\newcommand{\DP}{\mbox{$\Delta\phi$}}
\newcommand{\DR}{\mbox{$\Delta R$}}
\newcommand{\DE}{\mbox{$\Delta\eta$}}
\newcommand{\DEP}{\mbox{$\Delta\eta_{c}$}}
\newcommand{\DEC}{\mbox{$\Delta\eta_{c}$}}
\newcommand{\SP}{\mbox{$S(\DEP)$}}
\newcommand{\PH}{\mbox{$\phi$}}
\newcommand{\EA}{\mbox{$\eta$}}
\newcommand{\EAJ}{\mbox{\EA(jet)}}
\newcommand{\AEA}{\mbox{$|\eta|$}}
\newcommand{\Ge}[1]{\mbox{#1 GeV}}
\newcommand{\T}[1]{\mbox{#1 TeV}}
\newcommand{\D}[1]{\mbox{$#1^{\circ}$}}
\newcommand{\x}{\cdot}
\newcommand{\ra}{\rightarrow}
\newcommand{\mb}{\mbox{mb}}
\newcommand{\nb}{\mbox{nb}}
\newcommand{\ipb}{\mbox{${\rm pb}^{-1}$}}
\newcommand{\inb}{\mbox{${\rm nb}^{-1}$}}
\newcommand{\rs}{\mbox{$\sqrt{s}$}}
\newcommand{\fdel}{\mbox{$f(\DEP)$}}
\newcommand{\fdele}{\mbox{$f(\DEP)^{exp}$}}
\newcommand{\fgap}{\mbox{$f(\DEP\! > \!3)$}}
\newcommand{\fgape}{\mbox{$f(\DEP\! > \!3)^{exp}$}}
\newcommand{\fpyt}{\mbox{$f(\DEP\!>\!2)$}}
\newcommand{\delth}{\mbox{$\DEP\! > \!3$}}
\newcommand{\uplim}{\mbox{$1.1\!\times\!10^{-2}$}}
\newcommand{\sigew}{\mbox{$\sigma_{\rm EW}$}}
\newcommand{\sigsi}{\mbox{$\sigma_{\rm singlet}$}}
\newcommand{\sigr}{\mbox{$\sigsi/\sigma$}}
\newcommand{\sigrew}{\mbox{$\sigew/\sigma$}}
\newcommand{\ncal}{\mbox{$n_{\rm cal}$}}
\newcommand{\ntrk}{\mbox{$n_{\rm trk}$}}

\def\simge
{\mathrel{\rlap{\raise 0.53ex \hbox{$>$}}{\lower 0.53ex \hbox{$\sim$}}}}

\def\simle
{\mathrel{\rlap{\raise 0.4ex \hbox{$<$}}{\lower 0.72ex \hbox{$\sim$}}}}

%
\def\sigtot{$\sigma_{\rm tot}$}         
\def\sigtop{$\sigma_{t \overline{t}}$}  
\def\pbarp{$\overline{p}p $}            
\def\ppbar{$p\overline{p} $}            
\def\qqbar{$q\overline{q}$}             
\def\ttbar{$t\overline{t}$}             
\def\bbbar{$b\overline{b}$}             
\def\D0{D\O}                            
\def\ipb{pb$^{-1}$}                     
\def\pt{p_T}                            
\def\ptg{p_T^\gamma}                    
\def\et{E_T}                            
\def\etg{E_T^\gamma}                    
\def\htran{$H_T$}                       
\def\gevcc{{\rm GeV}/c$^2$}                   
\def\gevc{{\rm GeV}/c}                  
\def\gev{\rm GeV}                       
\def\tev{\rm TeV}                       
\def\njet{$N_{\rm jet}$}                
\def\aplan{$\cal{A}$}                   
\def\lum{$\cal{L}$}                     
\def\iso{$\cal{I}$}                     
\def\remu{${\cal{R}}_{e\mu}$}           
\def\rmu{$\Delta\cal{R}_{\mu}$}         
\def\pbar{$\overline{p}$}               
\def\tbar{$\overline{t}$}               
\def\bbar{$\overline{b}$}               
\def\lumint{$\int {\cal{L}} dt$}        
\def\lumunits{cm$^{-2}$s$^{-1}$}        
\def\etal{{\sl et al.}}                 
\def\vs{{\sl vs.}}                      
\def\sinthw{sin$^2 \theta_W$}           
\def\mt{$m_t$}                          
\def\mb{$m_b$}                          
\def\mw{$M_W$}                          
\def\mz{$M_Z$}                          
\def\pizero{$\pi^0$}                    
\def\jpsi{$J/\psi$}                     
\def\wino{$\widetilde W$}               
\def\zino{$\widetilde Z$}               
\def\squark{$\widetilde q$}             
\def\gluino{$\widetilde g$}             
\def\alphas{$\alpha_{\scriptscriptstyle S}$}                
\def\alphaem{$\alpha_{\scriptscriptstyle{\rm EM}}$}         
\def\epm{$e^+e^-$}                      
\def\etg{\mbox{$E_T^\gamma$}}           
\def\etag{\mbox{$\eta^\gamma$}}         
\def\met{\mbox{${\hbox{$E$\kern-0.6em\lower-.1ex\hbox{/}}}_T$}} 
\newcommand{\NC}{{\em Nuovo Cimento\/} }
\newcommand{\NIM}{{\em Nucl. Instr. Meth.} }
\newcommand{\NP}{{\em Nucl. Phys.} }
\newcommand{\PL}{{\em Phys. Lett.} }
\newcommand{\PR}{{\em Phys. Rev.} }
\newcommand{\PRL}{{\em Phys. Rev. Lett.} }
\newcommand{\RMP}{{\em Rev. Mod. Phys.} }
\newcommand{\ZP}{{\em Zeit. Phys.} }
\def\err#1#2#3 {{\it Erratum} {\bf#1},{\ #2} (19#3)}
\def\ib#1#2#3 {{\it ibid.} {\bf#1},{\ #2} (19#3)}
\def\nc#1#2#3 {Nuovo Cim. {\bf#1} ,#2(19#3)}
\def\nim#1#2#3 {Nucl. Instr. Meth. {\bf#1},{\ #2} (19#3)}
\def\np#1#2#3 {Nucl. Phys. {\bf#1},{\ #2} (19#3)}
\def\pl#1#2#3 {Phys. Lett. {\bf#1},{\ #2} (19#3)}
\def\prev#1#2#3 {Phys. Rev. {\bf#1},{\ #2} (19#3)}
\def\prl#1#2#3 {Phys. Rev. Lett. {\bf#1},{\ #2} (19#3)}
\def\rmp#1#2#3 {Rev. Mod. Phys. {\bf#1},{\ #2} (19#3)}
\def\zp#1#2#3 {Zeit. Phys. {\bf#1},{\ #2} (19#3)}


\title{
\begin{flushright}
{\normalsize FERMILAB-Pub-96/446-E}\\
{\normalsize 12 December 1996}
\end{flushright}
\vspace{-0.25cm}
Search for Diphoton Events with Large
Missing Transverse Energy\\
in $p \overline p$ Collisions at $\sqrt{s}=1.8\,{\rm TeV}^*$
\vspace{-0.25cm}}

%
\author{                                                                        
S.~Abachi,$^{14}$                                                               
B.~Abbott,$^{28}$                                                               
M.~Abolins,$^{25}$                                                              
B.S.~Acharya,$^{43}$                                                            
I.~Adam,$^{12}$                                                                 
D.L.~Adams,$^{37}$                                                              
M.~Adams,$^{17}$                                                                
S.~Ahn,$^{14}$                                                                  
H.~Aihara,$^{22}$                                                               
G.~\'{A}lvarez,$^{18}$
% ...
\\                                                                              
\vskip -0.10cm                                                                   
\centerline{(D\O\ Collaboration)}                                               
\vskip 0.20cm                                                                  
}                                                                               
\address{                                                                       
\centerline{$^{1}$Universidad de los Andes, Bogot\'{a}, Colombia}               
\centerline{$^{2}$University of Arizona, Tucson, Arizona 85721}                 
\centerline{$^{3}$Boston University, Boston, Massachusetts 02215}               
\centerline{$^{4}$Brookhaven National Laboratory, Upton, New York 11973}        
\centerline{$^{5}$Brown University, Providence, Rhode Island 02912}             
\centerline{$^{6}$Universidad de Buenos Aires, Buenos Aires, Argentina}         
\centerline{$^{7}$University of California, Davis, California 95616}            
\centerline{$^{8}$University of California, Irvine, California 92717}           
\centerline{$^{9}$University of California, Riverside, California 92521}        
\centerline{$^{10}$LAFEX, Centro Brasileiro de Pesquisas F{\'\i}sicas,          
                  Rio de Janeiro, Brazil}                                       
\centerline{$^{11}$CINVESTAV, Mexico City, Mexico}                              
\centerline{$^{12}$Columbia University, New York, New York 10027}               
\centerline{$^{13}$Delhi University, Delhi, India 110007}                       
\centerline{$^{14}$Fermi National Accelerator Laboratory, Batavia,              
                   Illinois 60510}                                              
\centerline{$^{15}$Florida State University, Tallahassee, Florida 32306}        
\centerline{$^{16}$University of Hawaii, Honolulu, Hawaii 96822}                
\centerline{$^{17}$University of Illinois at Chicago, Chicago, Illinois 60607}  
\centerline{$^{18}$Indiana University, Bloomington, Indiana 47405}              
\centerline{$^{19}$Iowa State University, Ames, Iowa 50011}                     
\centerline{$^{20}$Korea University, Seoul, Korea}                              
\centerline{$^{21}$Kyungsung University, Pusan, Korea}                          
\centerline{$^{22}$Lawrence Berkeley National Laboratory and University of      
                   California, Berkeley, California 94720}                      
\centerline{$^{23}$University of Maryland, College Park, Maryland 20742}        
\centerline{$^{24}$University of Michigan, Ann Arbor, Michigan 48109}           
\centerline{$^{25}$Michigan State University, East Lansing, Michigan 48824}     
\centerline{$^{26}$Moscow State University, Moscow, Russia}                     
\centerline{$^{27}$University of Nebraska, Lincoln, Nebraska 68588}             
\centerline{$^{28}$New York University, New York, New York 10003}               
\centerline{$^{29}$Northeastern University, Boston, Massachusetts 02115}        
\centerline{$^{30}$Northern Illinois University, DeKalb, Illinois 60115}        
\centerline{$^{31}$Northwestern University, Evanston, Illinois 60208}           
\centerline{$^{32}$University of Notre Dame, Notre Dame, Indiana 46556}         
\centerline{$^{33}$University of Oklahoma, Norman, Oklahoma 73019}              
\centerline{$^{34}$University of Panjab, Chandigarh 16-00-14, India}            
\centerline{$^{35}$Institute for High Energy Physics, 142-284 Protvino, Russia} 
\centerline{$^{36}$Purdue University, West Lafayette, Indiana 47907}            
\centerline{$^{37}$Rice University, Houston, Texas 77005}                       
\centerline{$^{38}$Universidade Estadual do Rio de Janeiro, Brazil}             
\centerline{$^{39}$University of Rochester, Rochester, New York 14627}          
\centerline{$^{40}$CEA, DAPNIA/Service de Physique des Particules, CE-SACLAY,   
                   France}                                                      
\centerline{$^{41}$Seoul National University, Seoul, Korea}                     
\centerline{$^{42}$State University of New York, Stony Brook, New York 11794}   
\centerline{$^{43}$Tata Institute of Fundamental Research,                      
                   Colaba, Bombay 400005, India}                                
\centerline{$^{44}$University of Texas, Arlington, Texas 76019}                 
\centerline{$^{45}$Texas A\&M University, College Station, Texas 77843}         
}                                                                               

\maketitle

\vspace{0.1cm}
\begin{abstract}
A search for signals of new physics has been carried out
in the channel $p \overline p \to \gamma \gamma + \met$.
This signature is expected in
various recently proposed supersymmetric (SUSY) models.  
We observe 842 events with two photons having transverse momentum
$\etg > 12\,$GeV
and pseudorapidity $|\etag|<1.1$. Of these, none have missing transverse
energy ($\met$) in excess of 25~GeV.  
The distribution of $\met$ is consistent with that of the
expected background.
We therefore set limits on production cross sections for
selectron, sneutrino and neutralino pairs, decaying into photons.
The limits range from about 400~fb to 1~pb depending on the 
sparticle masses. A general limit of 185~fb (95\% C.L.)
is set on $\sigma\cdot B(p\overline p \to \gamma\gamma\met+X)$,
where $\etg > 12\,$GeV,  $|\etag|<1.1$, and $\met > 25\,$GeV.

\vspace{0.2cm}
\par\noindent{\hbox to 3cm{\hrulefill}}
\par\noindent $*$ Submitted to Physical Review Letters
\end{abstract}


\twocolumn



We have searched for new physics in the channel
$p \overline p \to \gamma \gamma\met+X$ (where $\met$ denotes missing
transverse energy). 
This was motivated by recent 
suggestions\cite{kane,dimopoulos,dine,kanetwo,kanethree,baer} 
that low-energy supersymmetry (SUSY)
may result in signatures involving one or more 
photons together with missing transverse energy.  The predicted cross
sections were suggested to be high enough to lead to 
several tens of events in present data, making signals easily 
detectable.  The recent theoretical analyses were motivated by a desire to 
explain a single $ee\gamma\gamma+\met$ event observed by the 
CDF collaboration\cite{cdf}. 

For this analysis, data corresponding to an integrated luminosity of
$93.3\pm 11.2\,{\rm pb}^{-1}$, recorded during 1992--95 with the
\D0\ detector\cite{dzero}, were used.  
Photons were identified using the uranium-liquid argon sampling calorimeter,
which covers the region of pseudorapidity $|\eta| = 
| -\ln\tan{\theta\over 2}|
\simle 4$. The electromagnetic (EM) energy 
resolution is
$\sigma_E/{E} = 15\%/\sqrt{E (\rm GeV)} \oplus 0.3\%$. The EM calorimeter
is segmented into four longitudinal sections, and transversely
into towers in pseudorapidity and azimuthal angle, of size 
$\Delta \eta \times \Delta \phi = 0.1 \times 0.1$ ($0.05 \times 0.05$ at shower
maximum).
Drift chambers in front of the calorimeter were used to distinguish
photons from electrons and photon conversions.
A three-level triggering system was employed.
The first level used scintillation counters near the 
beam pipe to detect an inelastic interaction;
the second level summed 
the EM energy in calorimeter towers of size
$\Delta \eta \times \Delta \phi =0.2 \times 0.2 $. 
The third level was a software 
trigger which formed clusters of calorimeter cells and applied
loose cuts on the shower shape. 

Events were selected which had two photon candidates, each with
transverse energy $E_T^\gamma > 12\,$GeV and $|\etag|<1.1$.
Each cluster was required to pass photon-selection requirements\cite{prl},
namely to have a shape consistent with that of a single EM
shower,
to have more than 96\% of its energy in the EM section
of the calorimeter, and to be 
isolated. The latter was based on the transverse energy $E_T^{\rm iso}$ in the
annular region between
${\cal R}=\sqrt{\Delta\eta^2 + \Delta\phi^2}=0.2$ and ${\cal R}=0.4$ 
around the cluster, requiring
$E_T^{\rm iso} < 2\,$GeV. 
Candidates were rejected if the cluster was near
an azimuthal module boundary, if there
was a track (or a significant number of drift-chamber hits) in a tracking road
$\Delta\theta\times\Delta\phi = 0.2\times 0.2$ between the cluster and the 
vertex, if
the invariant mass of the photon pair was between 
80 and 100~GeV/$c^2$ (to reject misidentified $Z\to ee$ events), or 
if the azimuthal angle between the two photons was less than
$90^\circ$ (to reduce the background from $W\gamma$ production and
radiative $W\to e\nu\gamma$ decays, with the electron misidentified as a 
photon).
The beam pipe of the Main Ring accelerator passes through the outermost 
layer of the calorimeter. Losses of accelerated particles from the Main Ring 
can lead to energy deposits in the calorimeter and thus to spurious 
missing transverse energy.  To eliminate this source of background the 
$\met$ was required to have an azimuthal separation 
$20^\circ < \Delta\phi < 160^\circ$ from the Main Ring.
To eliminate events where the $\met$ was due to mismeasured jet energy,
it was also required to have an azimuthal separation
$\Delta\phi < 160^\circ$ from either of the leading two 
jets (provided $E_T^{\rm jet}> 12$~GeV).

\begin{figure}[t]
\epsfxsize=8cm
\dofig{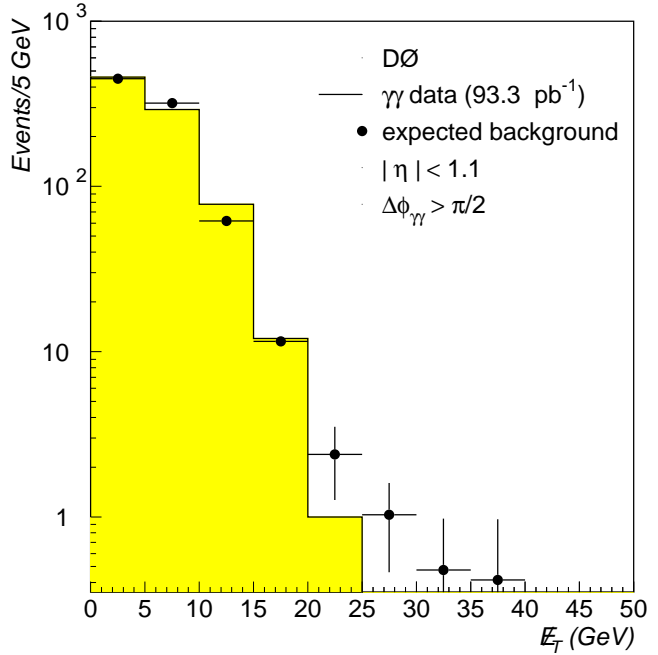}
\caption{Distribution of $\met$ 
for $\gamma\gamma$ data (shaded histogram), and for
the total expected background (black circles). 
\label{met}
}
\end{figure}

These selections yielded 842 events, whose $\met$ distribution is plotted
in Fig.~\ref{met}.  No events are observed with $\met > 25\,$GeV.
The resolution of the detector in $\met$ is about 
4~GeV for diphoton final states passing these kinematic selections.   

The dominant background to diphotons with large $\met$ arises
from QCD events where jet or vertex mismeasurement leads to excess $\met$.
Therefore, starting with the same trigger and dataset,
a background sample was selected which was expected to suffer from
the same mismeasurements.
Two EM clusters, satisfying the same kinematic and fiducial
cuts as the signal, were required. Both were required to 
have more than 90\% of their energy in the EM section of the calorimeter.
At least one of the two EM clusters was required to fail the strict photon 
isolation criterion ($E_T^{\rm iso}< 2$~GeV) but both were required
to have $E_T^{\rm iso}< 5$~GeV;
at least one of the clusters was required to have a bad shower shape, and
both were required to have either no track in
the road, or a track with a bad match to the cluster.
Electron backgrounds are evaluated separately.
The resulting sample was expected to
contain both QCD multijet events where two jets
fluctuated into highly-EM clusters, and the $\met$ was due to
mismeasurement; and QCD photon$+$jets events, where one photon was real and 
the other a fluctuated jet, and the $\met$ was again due to
mismeasurement.
This selection yielded 1678 events.  The distribution was normalized to the
$\gamma\gamma$ sample over the range $\met < 20$~GeV to estimate the
background at higher $\met$.  The resulting number of events expected with
$\met > 25$~GeV is $1.0\pm 0.7$. 

\begin{table}[t]
\begin{center}
\begin{tabbing}
\hbox to 8.5cm{\hrulefill}
\\
\=Process \quad\=Masses\=\ (GeV/\=c$^2$)\qquad\=\quad\qquad\=
$\sigma$ (95\% C.L.)\\
\>\>$\tilde e$\>$\tilde \nu$\>$\tilde\chi^0_2$\>$\tilde\chi^0_1$\>(pb)\\

\hbox to 8.5cm{\hrulefill}
\\
\>$\tilde e \tilde e$\>100\>---\>90\>50\> 0.715\\
\hbox to 8.5cm{\hrulefill}
\\
\>$\tilde\nu\tilde\nu$\>---\> 70\>50\>30\> 0.995\\
\>                \>---\> 70\>60\>30\> 0.805\\
\>                \>---\> 70\>60\>50\> ---\\
\>                \>---\> 80\>65\>55\> 21.6\\
\>                \>---\> 80\>70\>60\> 20.8\\
\>                \>---\> 90\>70\>65\> --- \\
\>                \>---\> 90\>80\>65\> 2.13\\
\>                \>---\> 90\>80\>70\> 54.7\\
\>                \>---\>100\>90\>70\> 0.765\\
\>                \>---\>100\>90\>80\> 4.65\\
\hbox to 8.5cm{\hrulefill}
\\
\>$\tilde\chi^0_2\tilde\chi^0_2$ \>---\>---\>60\>30\> 0.688\\
\>                 \>---\>---\>60\>30\> 0.715\\
\>                 \>---\>---\>60\>40\> 0.935\\
\>                 \>---\>---\>70\>30\> 0.555\\
\>                 \>---\>---\>70\>40\> 0.680\\
\>                 \>---\>---\>70\>50\> 1.03\\
\>                 \>---\>---\>80\>30\> 0.471\\
\>                 \>---\>---\>80\>40\> 0.610\\
\>                 \>---\>---\>80\>50\> 0.750\\
\>                 \>---\>---\>90\>40\> 0.424\\
\>                 \>---\>---\>90\>50\> 0.478
\end{tabbing}
\hbox to 8.5cm{\hrulefill}
\end{center}
\caption{Upper limits on pair production cross section
(95\% C.L.) obtained for each of the Monte Carlo
samples generated for this analysis, based on zero observed events.
(A dash in the limits column indicates that there was
insufficient acceptance for a limit to be set on this 
combination of masses.)}
\end{table}

Processes such as
$W\to e\nu$, $\tau \to e X$ and even $t\overline t \to e X$
contain genuine $\met$ and an electron whose track may be lost.
If these are combined with a real or fake photon, an
apparent $\gamma\gamma\met+X$ signal can result.
Again,
starting with the same trigger and dataset,
a sample of $e\gamma+X$ events was selected, having
two EM clusters satisfying the same kinematic and fiducial
cuts as the
signal; both of the clusters were required to pass the strict photon 
selection (isolation, shower shape, EM fraction);
one of the two clusters had to have exactly 
one drift chamber track in the road, with a good match 
to the cluster,
and the other cluster had to have no associated drift chamber hits or track.
These selections yielded 321 $e\gamma +X$ events.
To estimate 
the contribution of 
such events to the $\gamma\gamma\met+X$ signal, it is first necessary to
remove the QCD background component from the $e\gamma +X$ candidate sample.
This was done by normalizing these two distributions in the 
region of low missing transverse energy ($\met < 20\,$GeV), then 
subtracting the QCD distribution from that of the $e\gamma +X$ candidates.
The resulting distribution was then multiplied by the
ratio of probabilities for a genuine electron to be reconstructed as a photon
or as an electron, which is estimated (from $Z\to ee$ events) to be
$0.14\pm0.01$ for the selection criteria used here.  The resulting
$e\gamma +X$ contribution to the 
$\gamma\gamma\met+X$ sample is estimated to be
$1.1\pm0.1$ events.

\begin{figure}[t]
\epsfxsize=8cm
\dofig{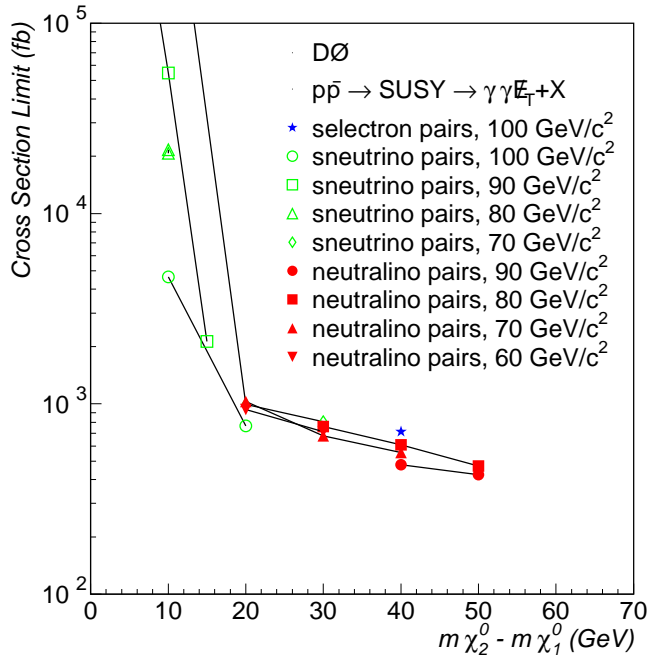}
\caption{Upper limits on pair production cross sections
(95\% C.L.) plotted as a function of neutralino mass 
difference.  The decay  $\tilde\chi^0_2 \to \gamma + \tilde\chi^0_1$ was
forced (see text).
\label{limit_plot}
}
\end{figure}

The total expected background is shown in
Fig.~\ref{met}, and agrees well with the observed data.  There is 
no evidence 
for non-standard sources of $\gamma\gamma$ events. The expected 
number of background 
events with $\met > 25$~GeV is $2.0\pm 0.9$ and none is observed. 
If we extend the pseudorapidity coverage for photons to $|\etag| < 2.5$,
we observe only one event with $\met > 30$~GeV, with an expected
background of $4.6 \pm 0.8$.

Simulated supersymmetry events were generated using
the {\tt ISAJET} Monte Carlo, version 7.20\cite{isajet}.  
The events were then processed through the detector simulation,
trigger simulation and the reconstruction software.
One thousand events were generated for each of the processes
and mass combinations listed in Table~I.
For the sneutrino-pair events, parameters were selected to keep the
chargino mass large enough so that the decay $\tilde \nu \to 
\tilde\chi^\pm_1 \ell$ 
remained kinematically forbidden.  
In all cases the decay $\tilde\chi^0_2 \to \tilde\chi^0_1 \gamma$ was forced.
The mean $\met$ and the mean photon $E_T$ in these events is 
typically $\sim m_{\tilde\chi^0_2}-m_{\tilde\chi^0_1}$, so we will primarily
be sensitive to cases where this mass difference exceeds about 20~GeV/$c^2$.  
Both photons are usually produced centrally, motivating
our requirement that $|\etag| < 1.1$.

The product of 
signal acceptance and efficiency, as estimated from these 
Monte Carlo samples,
is 
typically 0.05--0.10 for  
$m_{\tilde\chi^0_2} - m_{\tilde\chi^0_1} \simge 20$~GeV/$c^2$.
In addition to the Monte Carlo statistical error,  
a systematic uncertainty of 8\% has been included
(based on the level of agreement between Monte Carlo and
data-based estimates of the photon selection efficiencies).

Upper limits on the allowed cross sections for the
process $p\overline p \to \tilde x \tilde x$ were evaluated, based on
no events being observed for $\met > 25\,$ GeV. (This range of $\met$ is 
found to maximize the significance of the Monte Carlo supersymmetry
signals, given the observed background distribution.)   
Here $\tilde x =\tilde e, \tilde \nu, \tilde\chi^0_2$, with subsequent decays
$\tilde e, \tilde \nu \to
\tilde\chi^0_2$ and $\tilde\chi^0_2 \to \gamma + \tilde\chi^0_1$. 
No background contribution
was subtracted. 
The results are shown in Table I and Fig.~\ref{limit_plot}.  
The 95\% C.L. upper limits range from
about 400~fb to 1~pb for the cases 
with $m_{\tilde\chi^0_2} - m_{\tilde\chi^0_1} \geq 20$~GeV/$c^2$.

The results quoted above are somewhat model-dependent. They are
also difficult to relate to the light gravitino scenario 
of \cite{dimopoulos} and \cite{kanetwo}.
A general limit on final states with similar topologies
has therefore been derived. 
It is found that, provided $m_{\tilde\chi^0_2} - m_{\tilde\chi^0_1} \geq
20$~GeV$/c^2$, the acceptance for events with two photons
having $\etg > 12\,$GeV and  $|\etag|<1.1$, and with measured $\met > 25\,$GeV,
is independent of the production process ($\tilde e\tilde e, \tilde\nu
\tilde\nu, \tilde\chi^0_2 \tilde\chi^0_2$) and the sparticle 
masses (see Fig.~\ref{limit_plot}). 
The acceptance $\times$ efficiency is $0.183\pm 0.016$. (This includes a
diphoton acceptance and topological cut efficiency of 0.55,
an identification efficiency per photon of 0.75,
an azimuthal acceptance 
of 0.78 for the $\met$, and an efficiency of 0.79 for the $\met$ not to 
lie too close to a jet direction). 
The resulting limits are:
\begin{eqnarray*}
\sigma\cdot B(p\overline p \to \gamma\gamma\met+X) 
& < & 185\ {\rm fb}\ (95\%\ {\rm C.L.})\\
& < & 140\ {\rm fb}\ (90\%\ {\rm C.L.})
\end{eqnarray*}
where $\etg > 12\,{\rm GeV}$, $|\etag|<1.1,$ and $\met > 25\,{\rm GeV}$. 
These limits are stricter than those placed on the pair-production cross
sections (Table I) because typically only 25--50\% of the supersymmetry events
satisfy these kinematic requirements.
Comparison with Figs.~3 and 7 in Ref.~\cite{kanetwo} shows that this limit is
sufficient to rule out a large fraction of the proposed parameter space for
light gravitino models.  

In obtaining this limit we used our simulated supersymmetry events to estimate
the efficiency loss due to the relative azimuthal angle requirement
between the $\met$ and the
leading two jets ($E_T^j > 12\,$GeV).  
In the simulated events there were an average of 1.2 jets per event with
$E_T^j > 12\,$GeV. For final states with higher jet multiplicity
we would expect a small additional loss of efficiency ($\sim 10$\%) due to
the exclusion of additional azimuth. 

To summarize,
a search for signals of new physics has been carried out
in the channel $p \overline p \to \gamma \gamma \met +X$.
This signature is expected in
various recently proposed supersymmetric models.  
We observe 842 events with two photons having $\etg > 12\,$GeV
and $|\etag|<1.1$. Of these, none have $\met > 25\,$GeV.  
The distribution of $\met$ is consistent with that of the
expected background. 
We therefore set limits on production cross sections for
selectron, sneutrino and neutralino pairs decaying into photons
and non-interacting particles;
limits range from about 400~fb to 1~pb, depending on the 
sparticle masses. A general limit of 185~fb (95\% C.L.)
may also be set on $\sigma\cdot B(p\overline p \to \gamma\gamma\met+X)$
where $\etg > 12\,$GeV,  $|\etag|<1.1$, and $\met > 25\,$GeV. 
This is sufficient to exclude a considerable fraction of the parameter space
of recently proposed models.

%
We thank the staffs at Fermilab and the collaborating institutions for their
contributions to the success of this work, and acknowledge support from the 
Department of Energy and National Science Foundation (U.S.A.),  
Commissariat  \` a L'Energie Atomique (France), 
Ministries for Atomic Energy and Science and Technology Policy (Russia),
CNPq (Brazil),
Departments of Atomic Energy and Science and Education (India),
Colciencias (Colombia),
CONACyT (Mexico),
Ministry of Education and KOSEF (Korea),
CONICET and UBACyT (Argentina),
and the A.P. Sloan Foundation.

\end{document}